\documentclass[twocolumn]{aastex61}

\shorttitle{Oscillations in Sunspot's Superpenumbrae}
\shortauthors{A.~Chelpanov, N.~Kobanov}
\begin{document}
\title{Three-Minute Oscillations in Sunspot's Penumbrae and Superpenumbrae. Alfv\'{e}nic or Sound?}
\correspondingauthor{A.~\surname{Chelpanov}}
\email{chelpanov@iszf.irk.ru}
\author{Andrei~\surname{Chelpanov}}
\affil{Institute of Solar-Terrestrial Physics
                     of Siberian Branch of Russian Academy of Sciences, Irkutsk, Russia}
\author{Nikolai~\surname{Kobanov}}
\affiliation{Institute of Solar-Terrestrial Physics
                     of Siberian Branch of Russian Academy of Sciences, Irkutsk, Russia}

\begin{abstract}

In the immediate sunspots' vicinity---their superpenumbra---3-minute line-of-sight (LOS) velocity oscillations dominate in the photosphere and chromosphere.
Oscillations of similar periods are also registered in the transition region and lower corona above active regions.
The goal of the work is to clarify whether these LOS velocity oscillations are manifestations of Alfv\'enic waves in the lower solar atmosphere.
The study is based on three sunspots with the use of the Solar Dynamics Observatory data. Additional observations of a sunspot were carried out at ground-based Automated Solar Telescope.
We use narrow-band frequency filtration (5.6--5.8\,mHz) of the LOS velocity, magnetic field, and intensity signals of the Fe\,\textsc{i} 6173\,\AA\ spectral line.
For the analysis, we used 90-minute long series.
We come to the conclusion that the 3-minute oscillations in the LOS velocity signals result from magnetoacoustic waves rather than from Alfv\'enic waves.
However, oscillations registered in magnetic field signals indicate that Alfv\'enic waves may be present already in the photosphere.
Further research requires simultaneous observations of LOS velocity, magnetic field strength, spectral line width, and intensity carried out at two heights of the solar atmosphere.
 
\end{abstract}

\section{Introduction} \label{sec:intro}

Oscillations and waves in the solar atmosphere play a significant role in the processes of energy transportation. In the vast majority of studies, the propagation upwards is examined, from the photosphere through the chromosphere to the transition region, and higher to the corona. Sunspots are considered the most among the sources of oscillations. They provide a wide variety of physical characteristics, which presents considerable opportunities for studying oscillations.
From the first works that revealed running penumbral waves (RPWs) \citep{1969SoPh....7..351B, 1972ApJ...178L..85Z, 1972SoPh...27...71G} the number of studies that analyzed the details of these phenomena rapidly
grew \citep{1985A&A...143..201Z,1992ASIC..375..261L, 1995A&A...300..302Z, 1985ApJ...294..682L,1986ApJ...301.1005L, 2000A&A...363..306G, 2000SoPh..192..373B,2003SoPh..218...85B, 2006A&A...456..689T, 2006RSPTA.364..313B,2007A&A...473..943J, 2008A&A...481..811B, 2008ApJ...689.1379K,2010ApJ...722..131F, 2012ApJ...746..119R,2013ApJ...779..168J, 2015LRSP...12....6K, 2017ApJ...842...59J, 2018ApJ...852...15P, 2018ApJ...869..110S, 2019A&A...627A.169F, 2020NatAs...4..220J,2020ApJ...888...84S, 2021RSPTA.37900180S}.
In the course of the research, the understanding of the nature of the waves in sunspots has been transformed and refined.
For example, it was shown that the supposed horizontal propagation of RPWs with decreasing frequency is an apparent pattern resulting from different-frequency waves propagating along magnetic field lines with different inclinations \citep{2003A&A...403..277R, 2007ApJ...671.1005B, 2015ApJ...800..129M, 2015A&A...580A..53L, 2004A&A...424..671K, 2006SoPh..238..231K}. The issues related with studying RPWs are largely covered in a review \citet{2023LRSP...20....1J}.
Later, a more comprehensive picture of 3- and 5-minute oscillations in sunspots was developed that reaches up to coronal structures \citep{2012ApJ...757..160J,2015ApJ...812L..15K, 2023MNRAS.525.4815R,2024BSRSL..93..948R}.

Recently, studies on observations of the so-called Alfv\'enic waves were published \citep{2021RSPTA.37900183M, 2021ApJ...914L..16C, 2022ApJ...933..108C, 2023RvMPP...7...17M}. The term Alfv\'enic waves is used for the kink mode that, in contrast to pure Alfv\'en waves, is accompanied by intensity variations resulting from compression in plasma. Based on Doppler velocities in spectral data, \citet{2021ApJ...914L..16C} showed transverse movements caused by Alfv\'enic waves propagating horizontally in the image plane.

Our current work is motivated by \citet{2021ApJ...914L..16C, 2022ApJ...933..108C} and is dedicated to studying 3-minute oscillations observed in the superpenambral areas of active regions. We intend to describe the picture of the oscillations in the immediate vicinities of sunspots and their superpenumbrae, and study the signs that may determine the characteristics of the waves existing in the region. A particular question of interest is whether the 3-minute oscillations are mostly result from Alv\'{e}nic or sound MHD mode.

\section{Data and Analysis}

We start this section with a brief explanation of our approach to choosing objects for observations and analysis. Since most studies in the field pursue to yield comprehensive results relevant for all sunspots, one should, in our opinion, pick for such a research singular sunspots of a round symmetrical shape. Note that theoretical models of sunspots employ a round-shaped sunspot. Complex sunspot with a disintegrated umbra and penumbra may show many individual aspects that influence the analysis results.

We chose three sunspots for the analysis: NOAA 12794 observed on December 27, 2020 (hereafter Sunspot 1), NOAA 13111, October 3, 2022 (Sunspot 2), and NOAA 13140, November 10, 2022 (Sunspot 3). For each object, we used 90-minute series when a sunspot was at the central meridian.

The analysis is primarily based on the data from the Solar Dynamics Observatory \citep[SDO;][]{sdo}. The Atmospheric Imaging Assembly \citep[AIA;][]{sdoaia} onboard SDO provides series of full-disk images with a 0.6 arcsec/pixel spatial sampling and 12 second temporal sampling in a number of wavelength channels including the 304\,\AA\ and 171\,\AA\ channels. The Helioseismic and Magnetic Imager \citep[HMI;][]{sdohmi} provides Fe\,\textsc{i} 6173\,\AA\ intensity, LOS magnetic field and velocity for the full disk with a 0.5 arcsec/pixel sampling; its temporal resolution is 45 seconds.
\textbf{We prepared and aligned SDO data using the SolarSoft routines \textit{aia\underline{ }prep} and \textit{hmi\underline{ }prep}.
In order to de-rotate the data, we shifted the images in the East-West direction to achieve the highest correlation with the first image of the series.
Prior to this, we resized the images expanding them ten times to reduce the wobbling resulting from objects passing from one pixel to the next.}
We also used ground-based spectroscopic observations of active region NOAA\,11479 on May 16, 2012 (Spot\,4). Its observations were carried out using the Automated Solar Telescope at the Sayan Solar Observatory \citep{2015SoPh..290..363K} with a 3.5 second cadence. The guiding system of the telescope compensates the movement of the observed object due to the rotation of the Sun. The white-light images of the sunspots and the LOS magnetic field distribution in them are shown in Figure~1.

\begin{figure}
\centerline{
\includegraphics[width=8.5cm]{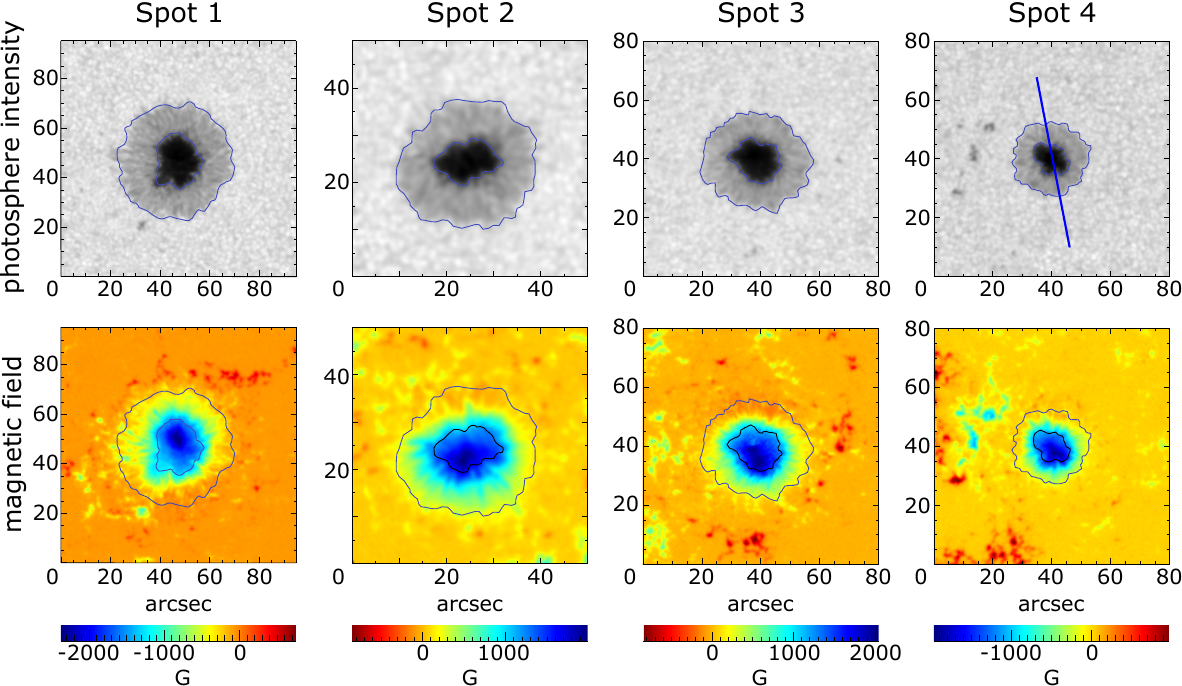}
}
\caption{The sunspots used in the analysis. \textit{Upper row}: the active regions in the photosphere in the HMI intensity channel; \textit{lower row}: distributions of the photospheric LOS magnetic field based of the HMI data. Sunspot\,4 was used in the AST observations. The location of the spectrograph slit is denoted in the top-right panel.}
\label{1}
\end{figure}

\section{Results}

For the purposes of the analysis, the umbra and penumbra boundaries are marked based on the HMI white-light images. The outer superpenubmra boundaries are not marked since their width typically exceeds 3--5 times that of penumbra \citep{1968SoPh....5..489L, 1974SoPh...38...77L, 2004ApJ...608.1148B}, so their borders lie outside the field of view in the figures. \citet{1974SoPh...38...77L} also divides superpenumbra into the inner and outer superpenumbra. In this classification, the inner superpenumbra in the chromosphere corresponds to the penumbra seen at the photospheric level. In this paper, however, we are using the term `penumbra' for both photospheric and chromospheric levels. We assume that superpenumbra starts immediately outside the penumbra outer boundary.

Figure~2 shows the maps of oscillation power in a given frequency range. The maps were constructed based on the Fourier oscillation power spectra calculated for each spatial point of the regions after the rotation compensation. The intensity of the colour in each point corresponds to the integrated Fourier power in the chosen frequency range.
To estimate the statistical significance, we used the procedure described in \citet{1998BAMS...79...61T} for the FFT power spectra normalized to 1/$\sigma^2$, where $\sigma^2$ is the variance of a time series. Throughout the paper, the power of the analyzed oscillations within the frequency ranges of interest exceeds the noise level.
The photospheric and chromospheric distributions of 3-minute oscillations over the sunspots and their vicinities show areas where the oscillation power concentrates within penumbrae and superpenumbrae.
They are also present at the upper photosphere level based on the Si\,\textsc{i} 10827\,\AA\ line Doppler velocity observations (Figure~3).

\begin{figure}
\centerline{
\includegraphics[width=8.5cm]{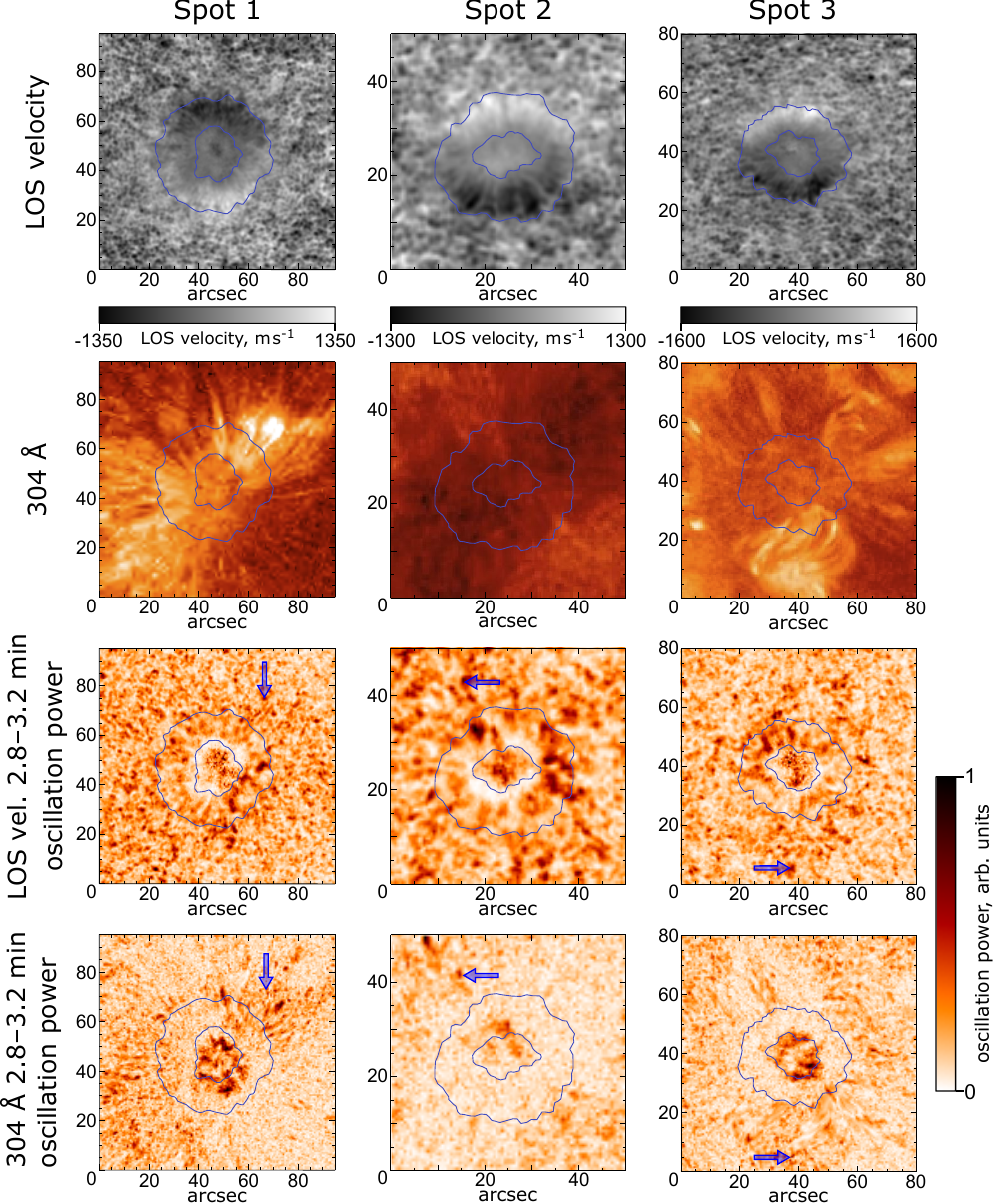}
}
\caption{Images of the sunspot regions in the the HMI photospheric Doppler velocity and AIA 304\,\AA\ channel signals and distributions of 3-minute (2.8--3.2\,minute range) oscillations in them. The arrows in the power distribution panels indicate examples of the locations, where 3-minute oscillation power is present in the superpenumbral areas in both photospheric velocity and 304\,\AA\ channels.}
\label{2}
\end{figure}

\begin{figure}
\centerline{
\includegraphics[width=4.5cm]{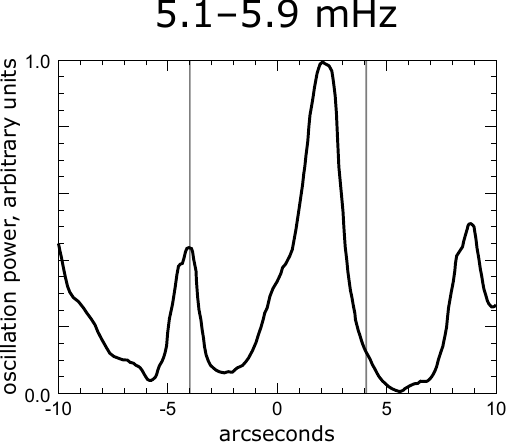}
}
\caption{Distribution of 3-minute oscillation power in the LOS-velocity signals in the NOAA\,11479 sunspot along the spectrograph's slit in the ground-based telescope in the Si\,\textsc{i} 10827\,\AA\ photospheric line. In addition to the main power peak in the central part of the umbra, one can see increased power regions in the inner and outer penumbra.}
\label{3}
\end{figure}

Such locations of increased 3-minute power oscillations are also seen in the 304\,\AA\ channel, whose formation height is attributed to the upper chromosphere and transition region \citep{2012ApJ...756...35R, 2016SoPh..291.3339K}. In some cases, these 3-minute power locations coincide with those in the photosphere maps (see arrow marks in the lower panels of Figure~2).
In most cases, however, the oscillation power locations are not in accordance between the atmosphere layers due to the inclination of the magnetic field.

In order to analyze the type of waves observed in these locations of penumbrae and superpenubrae, we compared the phases of the oscillations in the intensity and LOS-velocity signals. The intensity and LOS-velocity data are available in the SDO/HMI observations for the photospheric Fe\,\textsc{i} 6173\,\AA\ line. For sound waves, it is considered that the phase difference of around 180$^\circ$ indicates a propagating wave, and the difference of 90$^\circ$ indicates a standing wave \citep{1963ApJ...138..631N,1982ApJ...258..393L, 2011SSRv..158..397W}.
For the comparison, we used 10 to 12 points in each of the three sunspots located in umbrae and superpenumbrae.
The signals were filtered in a narrow range of periods centered at 3 minutes (2.8--3.2\,minutes) with the use of the wavelet algorithm. For this purpose we used the Interactive Data Language (IDL) wavelet software on the basis of the sixth-order Morlet wavelet developed and described by \citet{1998BAMS...79...61T}.
To measure the phase difference, we incrementally shifted the individual wave trains one period back and forth relative to each other to find the shift that would yield the highest correlation between the wave trains.
Figure~4 shows examples of intensity and velocity signals for several locations in Sunspot\,3. 
The comparison showed that for the majority of the locations in penumbrae, the phase shift during the series was close to 180 degrees, while in the superpenumbrae the phase difference was around 90$^\circ$ in approximately half of the locations, and 180$^\circ$ in the other half of the locations.
We also compared intensity and LOS velocity signals from the upper-photosphere Si\,\textsc{i} 10827\,\AA\ line available in our ground-based observations. As in the SDO data, the locations of heightened 3-minute oscillations in the penumbra show the phase difference close to 180$^\circ$. In the superpenumbra, the results are less consistent with no prevailing shift figures. Note, although, that in these data much fewer spatial locations are available, since in spectral observations the spectrograph's slit covers merely a narrow line of the solar surface.
The measured phase differences suggest that at the photospheric heights, the penumbra is mostly occupied with propagating waves, and in the superpenumbrae, the observed waves are a mixture of standing and propagating waves.

\begin{figure}
\centerline{
\includegraphics[width=8cm]{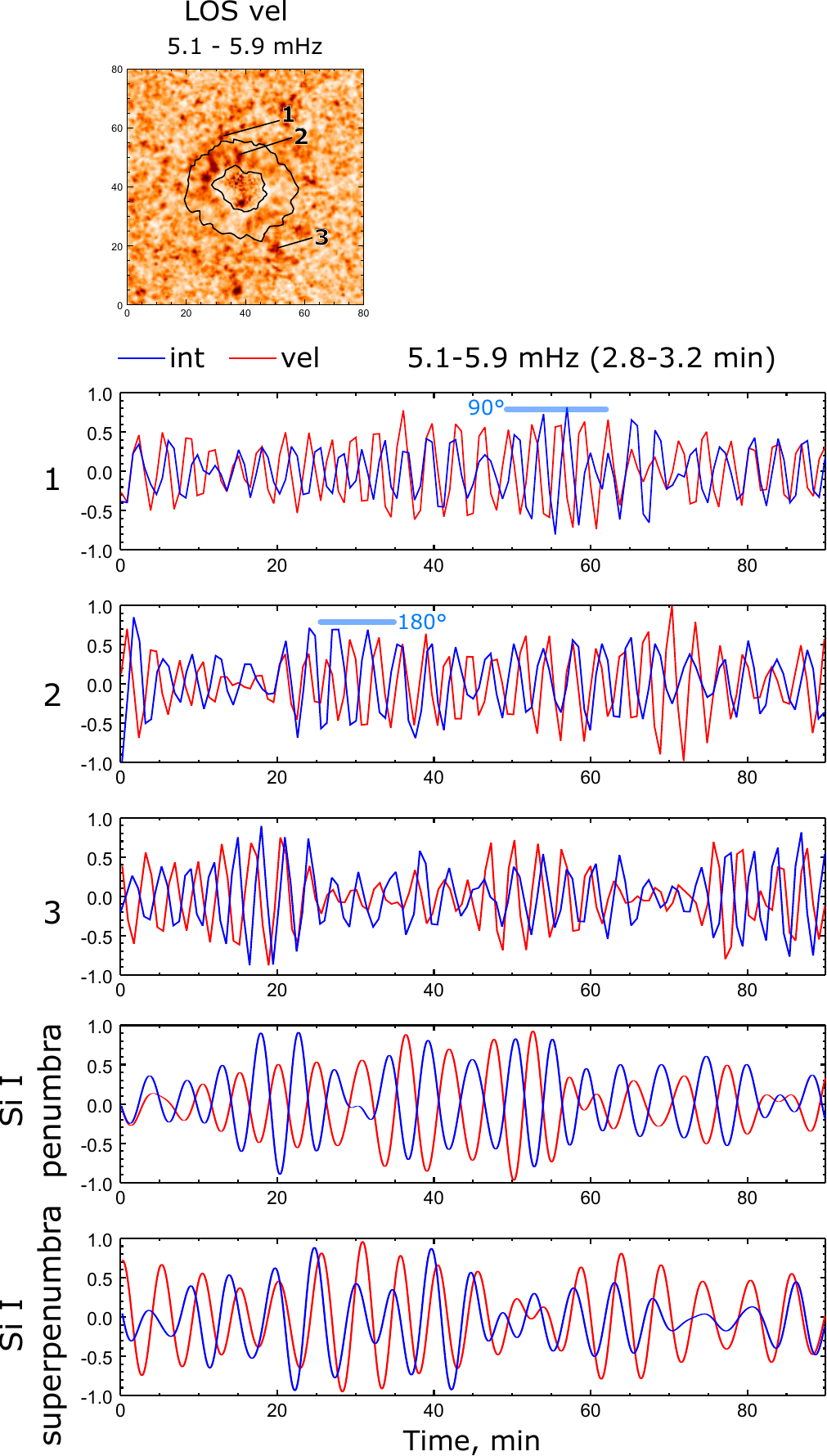}
}
\caption{Examples of phase comparisons between the intensity (blue) and LOS velocity (red) signals in the sunspot’s penumbra and superpenumbra in Sunspot\,3 (panels 1--3) and Sunspot\,4 (Si\,\textsc{i} 10827\,\AA\ line panels). Blue bars in oscillation panels 1 and 2 show examples when the phase difference between the signals is around 90$^\circ$ and 180$^\circ$.}
\label{4}
\end{figure}

Higher, in the 304\,\AA\ channel intensity at the upper-chromosphere level, 3-minute oscillations are observed above umbra and umbra-penumbra boundary. There are also areas of 3-minute oscillation power in the superpenumbra, although at this level they are less pronounced compared with those seen in the photospheric velocities
(Figure~2; examples of 3-minute oscillation power locations in the 304\,\AA\ panels marked with arrows).

In some cases, we also observed 3-minute oscillations in the photospheric LOS magnetic field signals based on the SDO/HMI data analysis. The LOS magnetic field dominant frequency maps in sunspots (Figure~5) show that most of the penumbra and superpenumbra is filled with noise (red background),
since the magnetic field oscillations are mostly quite weak, and their power is often below the noise level.
These maps are based on the Fourier spectra of the signals. For each point, a frequency is considered dominant if the integrated oscillation power in the 0.8\,mHz window centered at this frequency is maximal within the 1--8\,mHz range.
Along the superpenumbra perimeter, however, at a distance of 10--15 arcsec from the outer penumbra border, there are separate areas with 3.0--5.5\,mHz oscillations, whose power is above the noise level. These frequencies are close to those typically observed in the velocity and intensity signals. Interestingly, there are time intervals, when magnetic field oscillations are not accompanied by velocity or intensity oscillations, but exist autonomously from them. An example of such a location is given in Figure~5 (arrows in panels \textit{a} and \textit{b}). The magnetic field, photospheric intensity, and LOS-velocity signals filtered in the 2.8--3.2\,minute range are shown in panels \textit{c--e}. This point is located under a coronal loop seen in the 171\,\AA\ channel. The 171\,\AA\ intensity signal is also given in panel \textit{f}.
The autonomy of the oscillations in the magnetic field signals from those in the velocity and intensity signals lowers the probability of the magnetic field signals being a result of the velocity or intensity signals directly leaking into the magnetic field signal.

The 171\,\AA\ oscillations seem to not concur with those in the lower layers. \citep{2023MNRAS.525.4815R} showed that 3-minute oscillations observed in coronal loop structures in sunspot regions result from magnetoacoustic waves propagating from the photosphere of the umbra. Thus, the oscillation processes in the lower corona above superpenumbra have their roots in the central part of the sunspot as opposed to the lower atmosphere oscillations, that form entirely by the processes outside the sunspot.

\begin{figure}
\centerline{
\includegraphics[width=8.5cm]{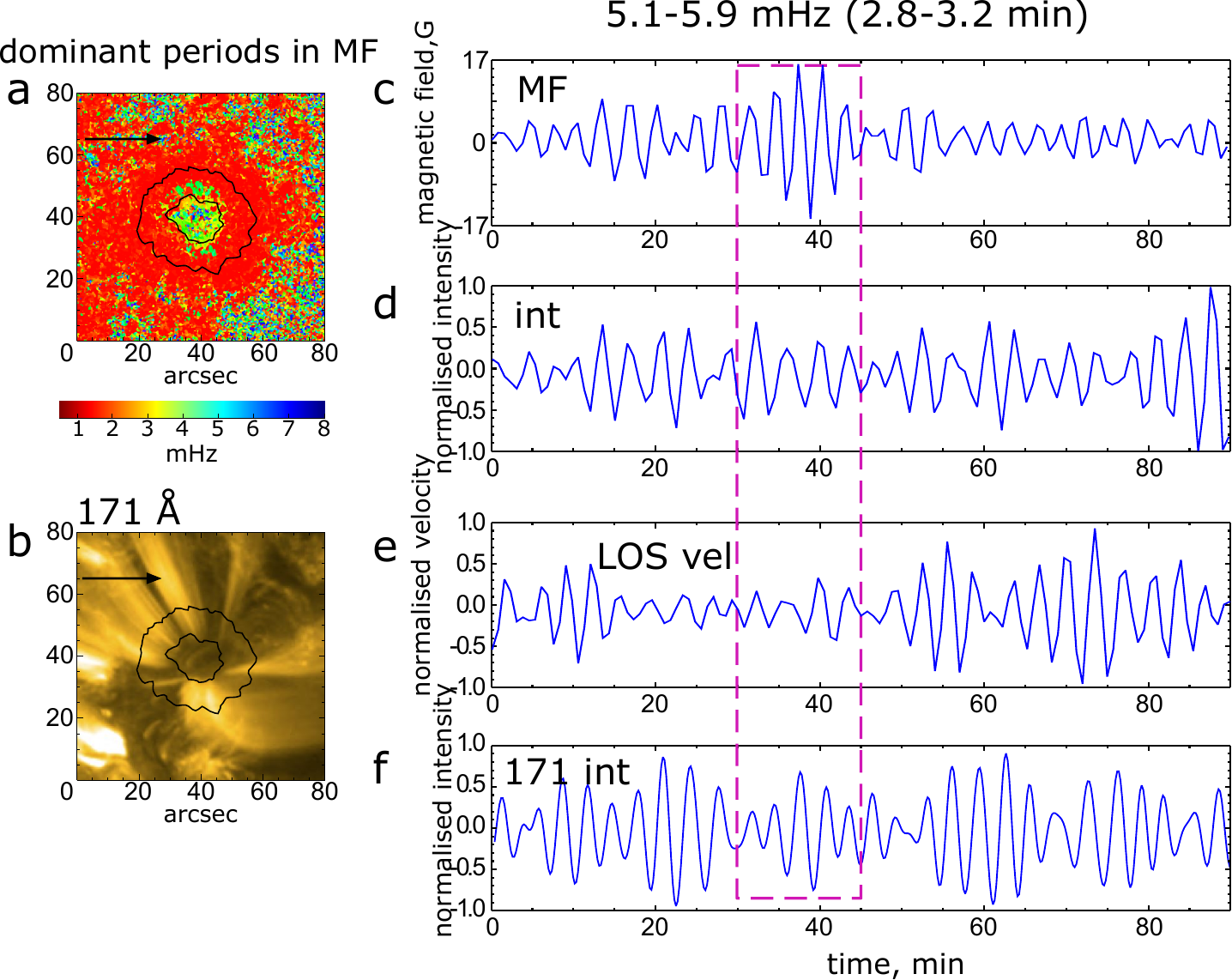}
}
\caption{\textit{a)} distribution of the dominant periods in the LOS magnetic field signals for Spot\,3; \textit{b)} the 171\,\AA\ image of the active region; the arrow in panels \textit{a} and \textit{b} show the location, where the signals are taken; \textit{c--f)} LOS magnetic field, photospheric intensity, LOS velocity, and 171\,\AA\ intensity signals filtered in the 3-minute range. The purple rectangle shows the time interval of the isolated LOS magnetic field wave train.}
\label{5}
\end{figure}

\section{Discussion}

The observation results show that locations of oscillation power of a 3-minute period exist in outer penumbrae and superpenumbrae. These areas are observed at a considerable height span between the lower photosphere and upper chromosphere. What input do these oscillations have in the general picture of the oscillations in a sunspot’s penumbra and superpenumbra? Can these areas stretched upwards be considered as channels that guide Alfv\'enic waves to the upper chromosphere and even to the corona?

Three-minute oscillations at the photospheric level in the superpenumbra show signs of propagating and standing waves depending on the location in roughly same proportions, as opposed to the penumbral regions, where mostly propagating wave signatures are observed.
This complies with the picture of the magnetic field topology in and around sunspots: magnetic field lines emerge in the penumbra and stretch outwards above the superpenumbra roughly horizontally.
Thus, in the penumbra we observe mostly propagating waves, while the superpenumbral regions show both propagating and standing wave signatures.

\textbf{\citet{2022ApJ...933..108C}} observed 3-minute oscillations and a secondary 10-minute period in H-alpha LOS velocity in the chromosphere along a superpenumbral fibril, which, according to the authors’ assessments, corresponded to horizontal magnetic field at the observed location. The authors conclude that these oscillations result from Alfv\'enic waves. In this configuration, LOS velocity oscillations orient perpendicular to the magnetic field lines, which led to the conclusion about Alfv\'enic waves. Based on their analysis of space-time diagrams, the authors estimated the wavelength of the observed wave to be as large as 18,000\,km. They proposed a model that may explain the conversion mechanism between the magnetoacoustic and Alfv\'enic modes.

The explanations that \textbf{\citet{2022ApJ...933..108C}} offer looks compelling, although, certain questions arise in regards of their interpretation of the oscillations as a result of Alfv\'enic waves.

These conclusions imply a homogeneous magnetic field configuration, where all the fibrils within the penumbra have the same inclination and similarly reach the horizontal position at the same distance from the sunspot. Such a configuration might justify the viewpoint that the periodic Doppler signals are exclusively due to the plasma moving perpendicularly to the magnetic field lines. High-resolution observations in sunspot penumbrae, however, have shown that the magnetic tube angle is uneven: more inclined tubes intersperse with less inclined forming the so-called uncombed configuration \citep{1991A&A...252..821D, 1992ASIC..375..195T, 1992A&A...264L..27S, 1993A&A...275..283S, 1993ASPC...46..173L, 2000A&A...361..734M}. One may assume that differently inclined tubes coexist in the superpenumbra as well. In such a case, Doppler signals from more vertical tubes would give an input to the measured LOS velocity signals in the penumbra, even though a significant part of the filaments in this region are nearly horizontal.

The existence in the superpenumbra of both locations with propagating and standing waves also supports the conclusion that the magnetic field in the region is unevenly inclined creating different conditions for wave propagation.


One can hardly agree with the assessment of the wavelength given in \citet{2021ApJ...914L..16C}.
The authors used the size of the patches in the images with no frequency filtering for 3-minute oscillations,
which are the main candidate for Alfv\'enic waves.
In this approach, different periods may contribute to the length of the observed tracks.

Thus, there are doubts whether these velocity oscillations serve as an unequivocal sign of Alfv\'enic waves. For a more definite conclusion, one should separate them from waves that possibly propagate upwards from the underlying layers, where similar periods are also registered. Regions with similar periods observed in the superpenumbra both in the photosphere and higher, for example, in the AIA 304\,\AA\ channel, may indicate the presence of such waves.

The phase comparison between the velocity and intensity signals shows that in the superpenumbra at the photospheric level, both standing and propagating waves populate different areas. This may indicate that inhomogeneous conditions for wave propagation exist in superpenumbra due to, for example, an uneven magnetic field configuration, whose direction and inclination play a role in the wave propagation behavior \citep{1977A&A....55..239B, 2016SoPh..291.3329C}.

In addition to the 3-minute oscillations, \textbf{\citet{2022ApJ...933..108C}} registered 10-minute oscillations in the LOS velocity signals.
Note that \citet{1994ASIC..433..197S} found that the Evershed flow was modulated with a 10-minute period. Earlier, \citet{2012ApJ...746..119R, 2013A&A...554A.146K} pointed that such long periods are observed in the outer penumbra at the photospheric and chromospheric levels.

To sum up, we assume that the oscillations in the observations of \textbf{\citet{2022ApJ...933..108C}} may as well not result from Alfv\'enic waves; in this case they are likely a manifestation of magnetoacoustic waves propagating between the atmosphere layers of the superpenumbra, since 3-minute oscillations in the lower atmosphere are predominantly attributed to the slow magnetoacoustic mode \citep{1972ApJ...178L..85Z,2006ApJ...640.1153C, 2015ApJ...812L..15K, 2016PhDT........15L}.
In this case, a question of no correlation between the velocity and intensity oscillations stays unexplained. The authors cite this lack of correlation as a proof that they observed virtually uncompressional oscillations.

Oscillations in the magnetic field signals may add to the picture of the wave-related processes in a sunspot region. Oscillations in the magnetic field signals unrelated to intensity and velocity oscillations are of particular interest. Different causes may potentially drive these oscillations. Among the possible explanations, there are periodic changes in the line formation height, which leads to the formation region residing consecutively in the more and less concentrated magnetic field \citep{2016SoPh..291.3329C}. Other than that, periodic movements of plasma along the solar surface may lead to a magnetic element successively entering and leaving the telescope’s aperture, which causes oscillations in the magnetic field signal. These movements, in turn, may result from the kink mode propagating along a vertical magnetic tube. Earlier, \citet{1971IAUS...43..348T} offered mechanisms that lead to oscillations in solar plasma being transferred to the LOS magnetic field signals.

Alfv\'en torsional waves may produce an oscillatory signal in magnetic field data. Torsional waves are propagating rotations of a magnetic tube that are not manifested in intensity and velocity signals. Twisting of a magnetic tube, though, may result in an increased magnetic field strength \citep{2010AIPC.1216..240M}. The torsional mode may also generate false variations in the magnetic field signal given the asymmetry of the spectral line. They also produce periodic variations in the spectral line width \citep{2009Sci...323.1582J,2022SoPh..297..154C}, which may translate into the measured magnetic field as apparent field oscillations.

Sausage mode is another type of MHD waves that may theoretically be responsible for variations in the LOS magnetic field signals. In this mode, periodical shrinking and expansion of a magnetic tube lead to an increase and decrease in the magnetic field strength. Sausage oscillations, however, also lead to density variations, hence intensity oscillations \citep{2011ApJ...729L..18M}. Thus, sausage oscillations should be manifested in both magnetic field and intensity signals. One should not rule out the crossover effect as a possible cause of oscillations in the magnetic field signals \citep{1972SoPh...22..119G, 1975SoPh...42...21G, 1974SoPh...37..113G}.

The question of the isolated variations in the magnetic field strength deserves a separate research.

\section{Conclusions}

In this work, we attempted to grasp the picture of the oscillations and waves typical of a sunspot superpenumbra---the closest vicinity of a sunspot, where magnetic field lines that emerge in the sunspot mostly reside horizontally and likely partially return under the surface. One of the main tasks is to assess which type of waves contributes the most into the observed oscillations: whether these are nearly non-compressive Alfv\'enic waves or sound waves propagating through the layers of the solar atmosphere. During the course of work, this proved to be a challenging task, and, in the future, it requires simultaneous observations of a number of parameters: LOS velocity, magnetic field strength, spectral line width, intensity, preferably, at two atmospheric heights. Nevertheless, we believe that the preliminary results that we obtain may attract interest of the readers.

Based on the SDO and ground-based data, sunspot superpenumbra is populated with patches of 3-minute oscillations, that are observed at a height range from the lower photosphere to the chromosphere.

Having analyzed the observations, we are leaning to the conclusion that we observe sound waves. A part of them is propagating waves, and another part is standing waves. We consider noteworthy but not convincing the assessment by \citet{2021ApJ...914L..16C} that the waves that they observed in the superpenumbra are Alfv\'enic waves.

A notable result in our observations is 3-minute oscillations in the LOS magnetic field that are not accompanied by oscillations in intensity or velocity signals. A number of factors may result in such oscillations that we discuss above, including Alfv\'enic waves propagating upwards. Establishing an exact nature of these oscillations requires additional studies involving simultaneous observations of LOS velocity, magnetic field strength, spectral line width, and intensity carried out at two or more heights of the solar atmosphere.

\acknowledgments

The work was financially supported by the Ministry of Science and Higher Education of the Russian Federation.
The spectral data were obtained using the equipment of Center for Common Use \textit{`Angara'} http://ckp-rf.ru/ckp/3056/ .

We thank the SDO science team for providing the data. We are grateful to an anonymous reviewer for their competent and useful suggestions that helped greatly improve the text.

\bibliography{Chelpanov}

\end{document}